\documentclass{article}
\usepackage{spconf,amsmath,graphicx,cite}

\usepackage{multirow}
\usepackage{amsxtra}
\usepackage{threeparttable}
\usepackage{amssymb,amsmath,bm}
\def\vec#1{\ensuremath{\bm{{#1}}}}

\title{Generative x-vectors for text-independent speaker verification}
\name{Longting Xu, Rohan Kumar Das, Emre Y\i lmaz, Jichen Yang and	Haizhou Li
	\thanks{This work is supported by the Neuromorphic Computing Programme under the RIE2020 Advanced Manufacturing and Engineering Programmatic Grant A1687b0033 in Singapore.}}
\address{Department of Electrical and Computer Engineering,\\
	National University of Singapore, Singapore\\
xltggn@gmail.com, rohankd@nus.edu.sg}

\begin{document}
	%
	\maketitle
\begin{abstract}
Speaker verification (SV) systems using deep neural network embeddings, so-called the x-vector systems, are becoming popular due to its good performance superior to the i-vector systems. The fusion of these systems provides improved performance benefiting both from the discriminatively trained x-vectors and generative i-vectors capturing distinct speaker characteristics. In this paper, we propose a novel method to include the complementary information of i-vector and x-vector, that is called generative x-vector. The generative x-vector utilizes a transformation model learned from the i-vector and x-vector representations of the background data. Canonical correlation analysis is applied to derive this transformation model, which is later used to transform the standard x-vectors of the enrollment and test segments to the corresponding generative x-vectors. The SV experiments performed on the NIST SRE 2010 dataset demonstrate that the system using generative x-vectors provides considerably better performance than the baseline i-vector and x-vector systems. Furthermore, the generative x-vectors outperform the fusion of i-vector and x-vector systems for long-duration utterances, while yielding comparable results for short-duration utterances.

\end{abstract}
\section{Introduction}
\label{sec:intro}
	
Speaker verification (SV) is to authenticate a person based on the voice samples~\cite{Kinnunen2010,hansen_review}. The factor analysis approaches for SV led to a new era with their achievement in having high performance~\cite{Kenny2005,jfa}. Later, the total variability model based i-vector system has been a benchmark for SV studies in the current decade~\cite{dehak2011front}. Recently, the deep neural network (DNN) based systems have been in the focus of the research community. Due to their good performance, DNN-based systems have been incorporated in most systems in the latest NIST SRE challenge~\cite{Lee2017,Torres2017,iitg_indigo}.
	
The initial attempts with DNNs for SV have been made in the context of i-vector speaker modeling in terms of computing the phonetic posteriors~\cite{dnnASR,dnn_odyssey_2014}. Alternative approaches extract bottleneck features from DNN acoustic models that are combined with the acoustic features~\cite{dnn_letter,mitchDNN}. However, such approaches require a large amount of transcribed data and may not be as effective for out-of-domain data ~\cite{dnn_letter}. This led to the exploration of end-to-end DNN systems for SV that learn the speaker models in a discriminative manner~\cite{Variani2014Deep,end2endICASSP,end2endSLT,emdend2endSLT,Li2017Deep}.
	
The recent work in this direction focuses on using the speaker embeddings that are scored with a probabilistic linear discriminant analysis (PLDA) based back-end~\cite{Snyder2017embedding,Brummer2018}. This kind of systems give comparable or better results to that obtained with i-vector speaker modeling. Further, they are proven to be very effective under short-duration utterance scenarios~\cite{Zhang2017End}. The study on the score level fusion of i-vector and embedding based systems~\cite{Snyder2017embedding} showed that the fused system outperforms the individual systems due to their complementary characteristics. Later on, the robustness of x-vectors has been explored by applying data augmentation~\cite{snyder2018x}.
	
Another strategy to improve the x-vector based SV is to include some input from the generative models as their fusion has been found promising. The embedding process is discriminative in nature, whereas the i-vector framework is a generative model. Specifically, x-vector extraction is achieved by training a DNN to discriminate among different output labels, while the i-vector model relies on a universal background model (UBM) to collect sufficient statistics for deriving speaker models. However, directly concatenating or score level fusion of these two models may not be an effective way for application-oriented systems as it increases the need of run-time computation and memory. This motivated us to develop an efficient way of including information from the generative model based on total variability modeling for an embedding based SV system. 
	
In this work, we propose a novel approach that learns a transformation matrix using the i-vectors and x-vectors from the background data to utilize both generative and discriminative characteristics. Canonical correlation analysis (CCA) between these vectors is used to derive this transformation model. The CCA has been used previously for analysis of correlation among different features~\cite{Das2016Exploring} and for fusion of multi-modal features in SV \cite{Sargin2006Multimodal}. Additionally, it has been used for co-whitening for short and long duration utterances in an i-vector system~\cite{lt2018}. In this work, the CCA is considered to maximize the correlation of the two models based on generative and discriminative paradigms to discover complementary attributes. The transformation model is then used to transform standard x-vectors, so that they also benefit from the input of generative model. Moreover, a comparison of the proposed system and the fusion of i-vector and x-vector systems is presented to highlight the impact of the work for practical systems. 

In the following sections, we first introduce the fundamentals of i-vector and x-vector approaches for SV in Section~\ref{secii}. Section~\ref{seciii} introduces the proposed framework of generative x-vectors. The results of the SV experiments using the proposed approach are reported in Section~\ref{seciv}. Finally, Section~\ref{conc} concludes the work.

\section{Speaker recognition paradigms: Generative vs. Discriminative}
\label{secii}
	
This section provides an explanation of the basics of i-vector and x-vector systems as they are studied for the proposed framework of generative x-vectors. The detailed structure with the parameters used for various modules of both the systems are also mentioned.

\subsection{The i-vector: a generative model}
\label{Sec:ivec}

An i-vector system is based on generative mode that is derived using total variability model (TVM)~\cite{dehak2011front}. The TVM is learned by unsupervised learning that is used to represent each utterance in a compact low-dimensional vector as follows
\begin{equation}
\vec{M}=\vec{m}+\mathbf{T}\vec{x}\label{eq1}
\end{equation}
where \(\vec{M}\) is Gaussian mixture model (GMM) mean supervector of an utterance, \(\vec{m}\) represents UBM mean supervector and total variability model \(\mathbf{T}\) to obtain the i-vector $\vec{x}$.

\subsection{The x-vector: a discriminative model}
\label{Sec:xvec}  

Generative models are successful due to the strong mathematical representations. However, considering the goal as speaker discrimination helps to increase the robustness. In this regard, researchers pay more attention on discriminative training for speaker recognition recently as discussed in the introduction. We consider x-vector as the discriminative baseline system, since it is comparable with i-vector systems for text-independent speaker recognition, especially for short utterances. The DNN embedding structure in our work basically follows the work of~\cite{Snyder2017embedding,snyder2018x}. We do not use any data augmentation in the current work, that deserves future exploration.

A time-delay neural network (TDNN)~\cite{Waibel1989} is trained using the same acoustic features as in the i-vector system. The TDNN model includes five frame-level hidden layers, all using rectified linear unit (ReLU) activation and batch normalization \cite{Ioffe2015}. The specific time-delay information of these frame-level layers are listed in Table \ref{tab:dnn}. A statistics pooling layer follows the output of the last frame-level layer which computes the mean and standard deviation of the frames of input segments. The mean and standard deviation are stacked in a manner such that the output dimension is doubled. The final two hidden layers are 512-dimensional pooling layers, also operating at segment level, prior to the softmax layer which targets speaker labels for each audio segment. The softmax and the second pooling layer are removed during the testing phase and 512-dimensional x-vectors are extracted at the output of the first pooling layer.

\begin{table}[t]
	\caption{\label{tab:dnn} {The time-delay configuration of the frame-level layers in the TDNN architecture}}
	\vspace{1mm}	
	\centering 
	\renewcommand{\arraystretch}{1.1}
	\begin{tabular}{c|c|c}			
		\hline	
		Layer index& Layer context & Output dimension\\ \hline
		1 &	(-2,-1,0,1,2)  &	512 \\ \hline
		2 &	(-2,0,2) &	512 \\ \hline
		3 &	(-3,0,3) &	512 \\ \hline
		4 &	0 &	512 \\ \hline
		5 &	0 &	1500 \\ \hline	     			     	
	\end{tabular}	
\end{table}  

\section{Generative x-vectors: DNN embeddings with generative model input}
\label{seciii}

In this work, we propose a novel approach to take the advantage of the correlation between the i-vectors and x-vectors to utilize their complementary nature of learning speaker models. A transformation model is learned using CCA by considering the i-vectors and the corresponding x-vectors as the input pairs from the background speech data. During the enrollment and the testing phase, the i-vector system is excluded from the pipeline and only x-vector system is considered whose output is linearly transformed using the transformation matrix obtained from the CCA model. We refer this transformed output as generative x-vector, henceforth referred to as x\(_\text{g}\)-vector, since it captures certain properties of the input generative model (i-vector model) during the transformation.

\begin{figure}[t]
	\centering
	\includegraphics[width=87mm]{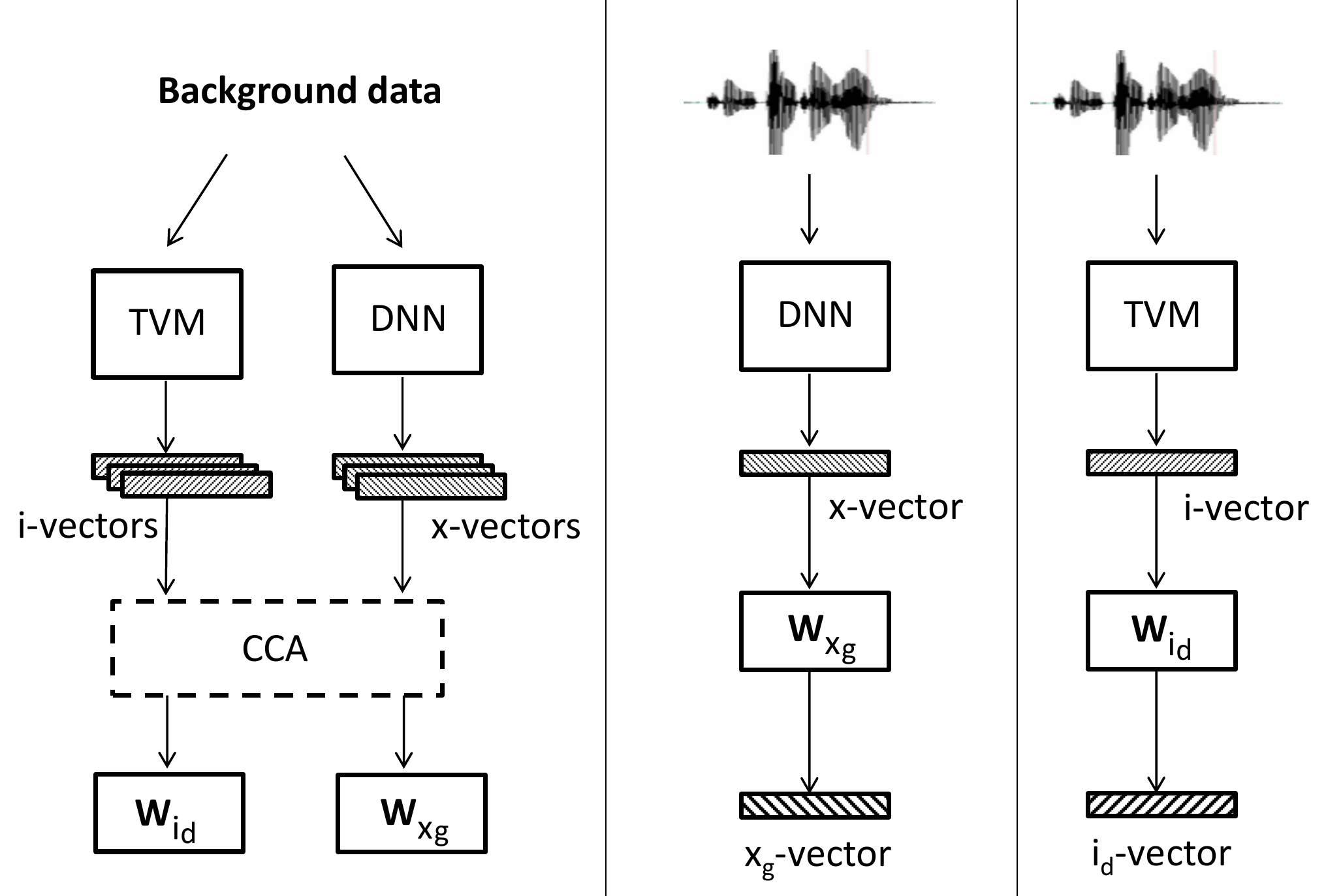} 	
	\caption{An overview of the proposed system, where the discriminative model (x-vector) benefits from the generative model (i-vector) input. The left panel shows the use of background data for CCA to train the transformation model \(\mathbf{W}_{x_\text{g}}\) and \(\mathbf{W}_{i_\text{d}}\). The middle panel shows computation of x\(_\text{g}\)-vector from x-vector using transformation model \(\mathbf{W}_{x_\text{g}}\). The right panel shows the contrast system of discriminative i-vector to generate i\(_\text{d}\)-vector using transformation model \(\mathbf{W}_{i_\text{d}}\).}
	\label{fig: diagram}
\end{figure}

Fig.~\ref{fig: diagram} illustrates the steps to obtain the proposed x\(_\text{g}\)-vector representation of speakers. During the training stage, a transformation matrix is learned by applying CCA and this matrix is later used for x\(_\text{g}\)-vector extraction. It is important to note that the TVM is only used for extracting i-vectors of the background data and is computed once. There is no further i-vector extraction involved during enrollment and test sessions. Hence, this kind of framework is expected to have relatively less latency than feature concatenation or score level fusion of these systems.

We first mathematically explain the left panel presented in Fig. \ref{fig: diagram}. In order to take advantage of the generative model information, we aim to seek a pair of matrices \(\mathbf{W}_{i_\text{d}}\) and \(\mathbf{W}_{x_\text{g}}\), which are confined in the following way
\begin{equation}
\max_{\mathbf{W}_{i_\text{d}},\mathbf{W}_{x_\text{g}}} corr(\mathbf{W}_{i_\text{d}}\mathbf{\Phi}_i,\mathbf{W}_{x_\text{g}}\mathbf{\Phi}_x) \label{equ:obj}
\end{equation}
Here, \(\mathbf{\Phi}_i\) and \(\mathbf{\Phi}_x\) contain the corresponding i-vectors and x-vectors from the same set of utterances. 

The proposed transformation with CCA is hypothesized to transfer information from the generative model to the discriminative model and vice-versa. Therefore, the resultant transformation matrices for i-vector and x-vector are denoted as \(\mathbf{W}_{i_\text{d}}\) and \(\mathbf{W}_{x_\text{g}}\), respectively. Let \(N\) be the number of background utterances used to train the transformation models with CCA. The dimension of background data i-vectors and x-vectors to CCA are \(N\times 600\)  and \(N\times 512\), respectively. On applying CCA, we obtain transformation matrices \(\mathbf{W}_{i_\text{d}}\) of size \(600\times 512\)  and  \(\mathbf{W}_{x_\text{g}}\) of size \(512\times 512\), respectively.

During the SV experiments, we only concentrate on the x-vector pipeline as shown in the right panel of Fig. \ref{fig: diagram}. Given an x-vector \(\phi_x\), the proposed vector is computed as
\begin{equation}
\phi_{x_\text{g}} = \mathbf{W}_{x_\text{g}}\phi_x
\end{equation}
where the \(\phi_{x_\text{g}}\) denotes the x-vector with generative model input that we refer to as a generative x-vector. 

Both the i-vectors and x-vectors are zero-centered in all of the mathematical expressions in this section. The details of CCA and the transformation of x-vectors are discussed in the following subsections.

\subsection{Canonical correlation analysis}
\label{Sec:CCA}

As mentioned in the aforementioned section, in this work, we aim to maximize the linear relationship between a set of i-vectors and x-vectors. It is to be mentioned that the dimensions for an i-vector and an x-vector are not the same. Given that a fixed number of background speech utterances is used to derive the background i-vectors and x-vectors, applying CCA maximizes the correlation between the input vector pairs of different dimensions.

Mathematically, given random vectors \(\vec{X}=(x_{1},\dots ,x_{n})^\mathrm{T}\) and \(\vec{Y}=(y_{1},\dots ,y_{m})^\mathrm{T}\), the CCA defines new set of variables \(U=\vec{a}^\mathrm{T}\vec{X}\) and \(V=\vec{b}^\mathrm{T}\vec{Y}\) via linear combinations of \(\vec{X}\) and \(\vec{Y}\)
\begin{equation}
U=\vec{a}^\mathrm{T}\vec{X}
\end{equation}
\begin{equation}
V=\vec{b}^\mathrm{T}\vec{Y}
\end{equation}
The CCA aims to find vectors \(\vec{a}\) and \(\vec{b}\) that maximizes the correlation \(\rho = {corr} \langle \vec{a}^\mathrm{T}\vec{X},\vec{b}^\mathrm{T}\vec{Y}\rangle \), which can written as  
\begin{equation}
\rho =\frac {E(\vec{a}^\mathrm{T}\vec{X}\vec{Y}^\mathrm{T}\vec{b})}{{\sqrt {E(\vec{a}^\mathrm{T}\vec{X}\vec{X}^\mathrm{T}\vec{a})}}{\sqrt {E(\vec{b}^\mathrm{T}\vec{Y}\vec{Y}^\mathrm{T}\vec{b})}}}
\end{equation} 
 With the constraints that
 \begin{equation}
\vec{a}^\mathrm{T}\mathbf{\Sigma}_{\vec{X}}\vec{a}=1
 \end{equation}
 and
 \begin{equation}
  \vec{b}^\mathrm{T}\mathbf{\Sigma}_{\vec{Y}}\vec{b}=1
 \end{equation}
 the correlation parameter to be maximized becomes 
\begin{equation}
\rho =\vec{a}^\mathrm{T}\mathbf{\Sigma}_{\vec{X}\vec{Y}}\vec{b} \label{eq:cca}
\end{equation}  
where \(\mathbf{\Sigma}_{\vec{X}}= E(\vec{X}\vec{X}^\mathrm{T}) \), \(\mathbf{\Sigma}_{\vec{Y}}= E(\vec{Y}\vec{Y}^\mathrm{T})\) and \(\mathbf{\Sigma}_{\vec{X}\vec{Y}}= E(\vec{X}\vec{Y}^\mathrm{T}) \) are the covariances.

We then obtain the first pair of canonical variates \((U_1,V_1)\) via maximizing \(\rho\) represented in Equation (\ref{eq:cca}). The remaining canonical variates \((U_l,V_l)\) maximize \(\rho\) subject to uncorrelated with \((U_k,V_k)\) for all \(k<l\). This procedure is iterated to \(\min\{m,n\}\) times that is based on the dimension of the two random vectors. Finally, we obtain \(\vec{a}_k\) is the \(k\)-th eigenvector of \(\mathbf{\Sigma} _{\vec{X}}^{-1}\mathbf{\Sigma} _{\vec{X}\vec{Y}}\mathbf{\Sigma} _{\vec{Y}}^{-1}\mathbf{\Sigma} _{\vec{Y}\vec{X}}\). Similarly, \(\vec{b}_k\) as the \(k\)-th eigenvector of \(\mathbf{\Sigma} _{\vec{Y}}^{-1}\mathbf{\Sigma} _{\vec{Y}\vec{X}}\mathbf{\Sigma} _{\vec{X}}^{-1}\mathbf{\Sigma} _{\vec{X}\vec{Y}}\). 

\begin{figure*}[t]
	\centering
	\includegraphics[width=150mm]{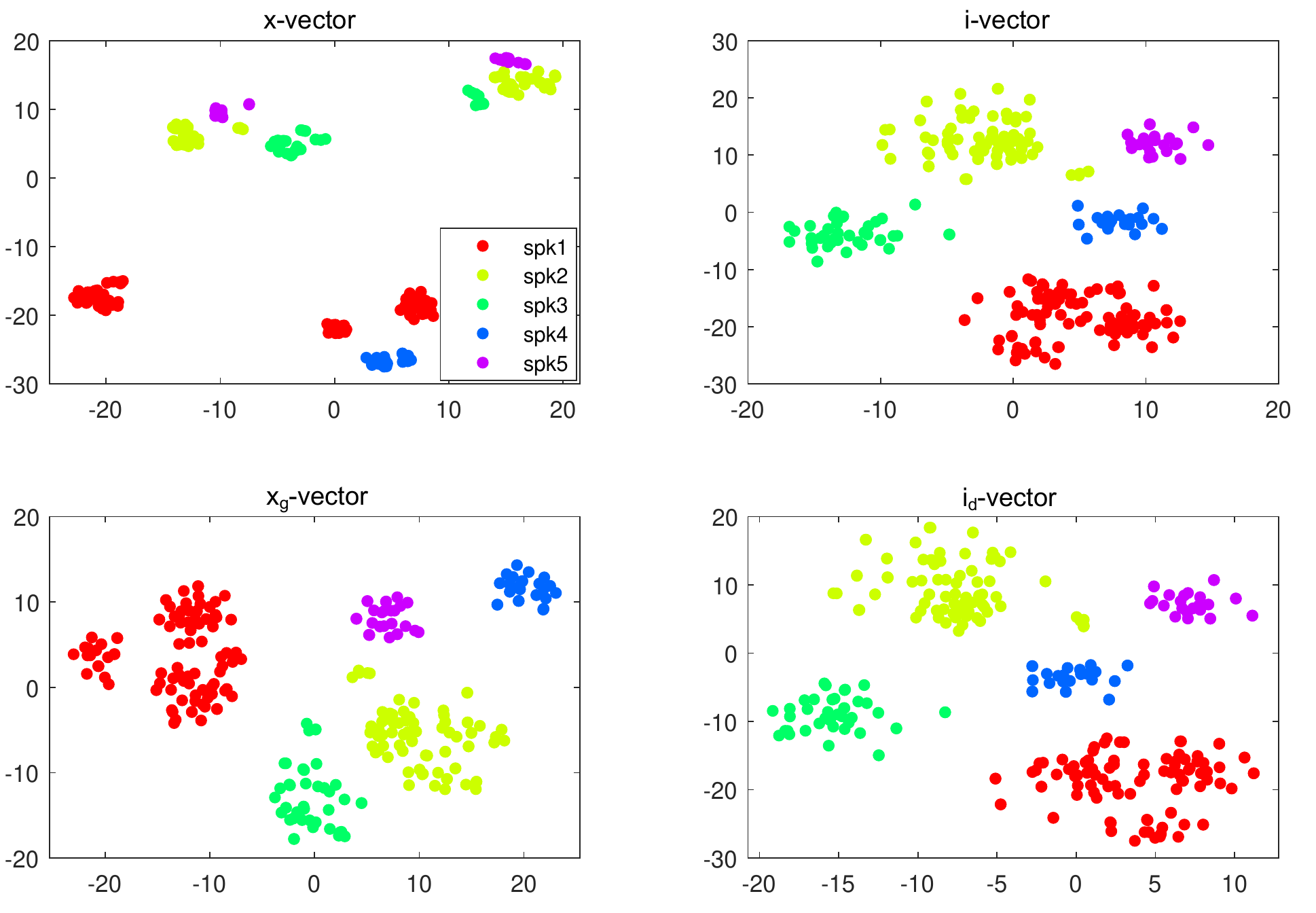} 	
	\caption{t-SNE visualization of different representations.}
	\label{fig:tsne}
   \vspace{-4mm}
\end{figure*}
\subsection{CCA based x-vector transformation}

In canonical correlation analysis we aim to find mutually orthogonal pairs of maximally correlated linear combinations of the variables in \vec{X} and \vec{Y}. In our work, the random vectors \vec{X} and \vec{Y} discussed in Section \ref{Sec:CCA} form the i-vector matrix \(\mathbf{\Phi}_i\) and x-vector matrix \(\mathbf{\Phi}_x\), respectively.

Revisiting the objective function given in Equation \eqref{equ:obj}, it can be solved with the following constraints,
\begin{equation}
\mathbf{W}_{i_\text{d}}\mathbf{\Sigma}_i\mathbf{W}_{i_\text{d}}^\mathrm{T} = \mathbf{I}
\end{equation}
and
\begin{equation}
\mathbf{W}_{x_\text{g}}\mathbf{\Sigma}_x\mathbf{W}_{x_\text{g}}^\mathrm{T} = \mathbf{I}
\end{equation}
where the x-vectors can be automatically whitened in the testing phase. Notice that \(\mathbf{\Sigma}_i\) and \(\mathbf{\Sigma}_x\) denote the empirical covariances of i-vectors and x-vectors, respectively. 

\subsection{t-SNE visualization}
\label{sec:sec_visual}

Together with the proposed  x\(_\text{g}\)-vector system, a contrast system is also introduced to derive another transformation model \(\mathbf{W}_{i_\text{d}}\) for the generative model i-vector to take input from the x-vector based discriminative model. The transformed i-vector is denoted by i\(_\text{d}\)-vector as input from discriminative model has been used. We visualize each speaker representation model to examine the distribution from subset of speakers using the t-Distributed Stochastic Neighbor Embedding (t-SNE) technique \cite{Maaten2008Visualizing}. The t-SNE technique is widely used for the visualization of high-dimensional data.

We have randomly chosen 5 speakers from the database that have more than 20 utterances and extracted corresponding i-vectors, x-vectors, x\(_\text{g}\)-vectors and i\(_\text{d}\)-vectors. Figure~\ref{fig:tsne} shows the t-SNE distributions for different representations. It is observed that the proposed x\(_\text{g}\)-vectors benefit from the generative information with an increased separability, while the distribution of i\(_\text{d}\)-vectors highly resembles the original i-vectors. 

The possible reason of this can be that the discriminative models like x-vectors learn the differences among the speakers without learning the characteristics of each speaker. Thus, when information from discriminative models is used as input to the generative model i-vector, it may not contribute towards a better SV performance. On the other hand, the generative models such as i-vectors, learn the characteristics of each speakers and they add specific speaker information when used as input to a discriminative model. Additionally, the discriminative models work well for a closed set of speakers, whereas there is no such constraint for generative models. 

\begin{table*}[t]
	\caption{\label{tab:comparison} {EER and DCF under CC'5 on NIST SRE 2010 database for different systems. Fusion results refer to score level fusion of i-vector and x-vector systems.}}
	\vspace{1mm}	
	\centering 
	\begin{tabular}{|c|c|c||c||c|c||c|c||c||c|c|}			
		\hline	
		\multirow{2}{*}{Tasks}&\multicolumn{5}{c||}{EER (\%)}& \multicolumn{5}{c|}{DCF}\\
		\cline{2-11}
		& i-vec & 	x-vec &fusion	& i\(_\text{d}\)-vec & 	x\(_\text{g}\)-vec& i-vec & 	x-vec &fusion	& i\(_\text{d}\)-vec & 	x\(_\text{g}\)-vec \\ \hline \hline
		coreext-coreext & 2.20 & 2.96 &
		2.19 &
		2.23 &
		1.51 &
		0.42 &
		0.42 &
		0.36 &
		0.44 &
		0.35
			\\ \hline	
		core-10sec &6.07 &
			6.39 &
		4.71 &
		6.00 &
		4.41 &
		0.85 &
			0.72 &
			0.78 &
		0.84 &
		0.70
			 \\ \hline
		10sec-10sec &11.46 &
		  11.51 &
		 8.92 &
	11.56 &
	8.93 &
	0.98 &
		 0.85 &
	0.88 &
	0.96 &
		0.89
		  \\ \hline		     			     	
	\end{tabular}				
 \vspace{2mm}
\end{table*}  

\section{Experimental Results}	
\label{seciv}

\subsection{Database}

The SV experiments in this work are performed using the NIST SRE 2010 database~\cite{nist2010}. The common condition 5 (CC'5) has been chosen for the evaluation. Further, we have considered different enrollment and test scenarios under this task, namely coreext-coreext, core-10sec, and 10sec-10sec, where coreext and core consist of long duration utterances, while 10sec denotes short-duration speech of 10 seconds. Additionally, Switchboard 2 Corpus of Phases 1, 2, and 3 as well as Switchboard Cellular, along with NIST SREs from 2004 to 2008 are considered as background data for learning the background models.

\subsection{Implementation details}

In this work, the 20-dimensional mel frequency cepstral coefficients (MFCC) features, along with delta and acceleration are extracted for each frame of 25 ms in shift of 10 ms. The i-vector model is used as a baseline system for reference in our studies. A full-covariance gender-independent UBM with 2048 components is used in the i-vector framework to obtain 600-dimensional i-vectors. For both the systems, dimensionality is reduced to 200 with linear discriminant analysis (LDA). For the x-vector system, the TDNN is trained on the same 20-dimensional MFCC features. All non-linearities in the neural network are ReLUs. 

We use PLDA for channel/session compensation and scoring in our experiments. Further, length normalization has been applied before performing PLDA~\cite{prince2007probabilistic}. The PLDA is trained to have 200 speaker factors with a full covariance, while the channel factor is ignored. The studies are reported in terms of equal error rate (EER) and detection cost function (DCF) that follows the protocol of NIST SRE 2010 evaluation plan~\cite{nist2010}. We used Kaldi recipes for building the baseline systems in this work~\cite{Povey_ASRU2011}.

\subsection{Results and discussion}
In this section, the results provided by the individual baseline systems using i-vectors and x-vectors are compared with the proposed generative x-vectors. We further apply score fusion to the i-vector and x-vector systems and compare with the generative x-vectors to investigate their effectiveness in capturing the complementary information from the generative model. 

Table \ref{tab:comparison} reports the performance of different SV frameworks used in this study. Comparing the i-vector and x-vector baselines, it is clear that the i-vector works better when both the enrollment and test utterances are of long durations, i.e., for coreext-coreext task. On the other hand, the results for core-10sec and 10sec-10sec tasks show that the x-vector system performs comparable to the i-vector system for short-duration test utterances when the enrollment data is either short or long. Further, a score level fusion of these two systems results in a gain for all considered tasks of the NIST SRE 2010 database. The system fusion results follow the trend reported by the authors of~\cite{Snyder2017embedding}.

We then focus on the results provided by the proposed x\(_\text{g}\)-vector system and its contrast i\(_\text{d}\)-vector system. It is observed that the proposed x\(_\text{g}\)-vector system outperforms the standard x-vector system by reducing the EER from 2.20\% to 1.51\%. On the other hand, the performance of the contrast i\(_\text{d}\)-vector system is similar to the original i-vector system. 
Finally, we compare the performance of proposed x\(_\text{g}\)-vector with the score level fusion. For short utterance cases, the performance of both systems are comparable. The proposed system outperforms the score fusion for the core condition with long utterances. Hence, the proposed system with a lower latency and less computational burden achieves a remarkable performance compared to fusion of the x-vector and i-vector systems. This highlights its importance as a field-deployable system in a practical setting.
\begin{figure}[t]
	\centering
	\includegraphics[width=80mm]{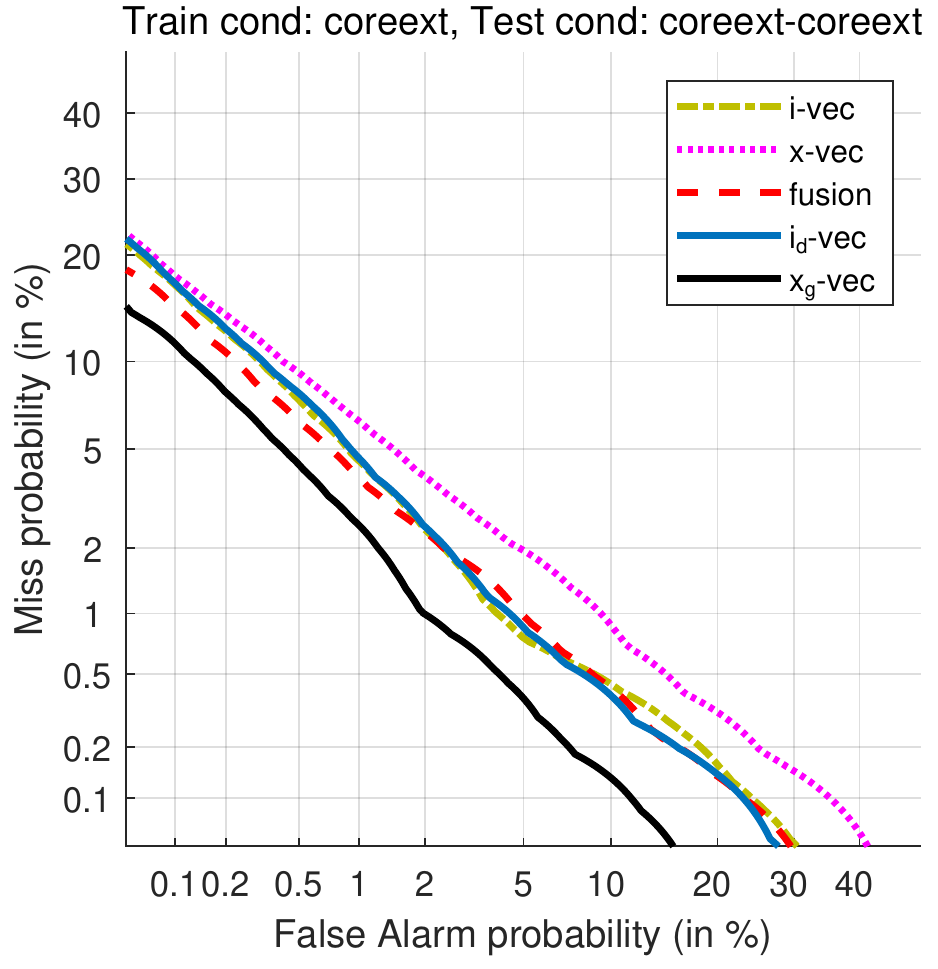} 	
	\caption{DET curves for the coreext-coreext task of NIST SRE 2010 database.}
	\label{fig:det}
	\vspace{-2mm}
\end{figure}

The detection error tradeoff (DET) curves for different systems obtained on the coreext-coreext task is illustrated in Fig. \ref{fig:det}. The superior performance of the proposed system is clearly reflected in this plot with a DET curve that is quite separate from the baseline individual systems as well as their fusion. In terms of EER, we observe 48.83\%, 22.46\% and 31.01\% relative improvement over the original x-vector system for the three different tasks of CC'5 on NIST SRE 2010 database discussed in this work.
The future work will focus on extending this framework with data augmentation to overcome mismatch conditions~\cite{snyder2018x,Mclaren2018,Novoselov2018}.

\section{Conclusions}
\label{conc}

This work focuses on having an improved DNN embedding based SV system that considers input from generative models. The total variability speaker modeling is used as the generative model for the studies. A transformation model is learned by applying CCA using background data i-vectors and x-vectors. This model is then used to obtain the generative x-vectors that are found to perform superior to its baseline as well as i-vector counterparts. The studies are performed on the NIST SRE 2010 database on three different conditions. The studies reveal 48.83\%, 22.46\% and 31.01\% relative improvement on EER for the coreext-coreext, core-10sec and 10sec-10sec tasks, respectively. This confirms the importance of using some inputs from the generative models for the framework of discriminative model of DNN embeddings for SV. Additionally, the performance of generative x-vectors is found to be superior for long utterances and competitive for short utterance cases to that obtained from the score level fusion of i-vector and x-vector systems. Thus, this kind of approaches have less latency than the dimension concatenation or score level fusion of systems that makes them useful for application purpose.  

\bibliographystyle{IEEE}
\bibliography{mybib}

\begin{thebibliography}{10}

\bibitem{Kinnunen2010}
T.~Kinnunen and H.~Li,
\newblock ``An overview of text-independent speaker recognition: From features
  to supervectors,''
\newblock {\em Speech Communication}, vol. 52, no. 1, pp. 12--40, 2010.

\bibitem{hansen_review}
J.~H.~L. Hansen and T.~Hasan,
\newblock ``Speaker recognition by machines and humans: A tutorial review,''
\newblock {\em IEEE Signal Processing Magazine}, vol. 32, no. 6, pp. 74--99,
  Nov 2015.

\bibitem{Kenny2005}
P.~Kenny,
\newblock ``Joint factor analysis of speaker and session variability: Theory
  and algorithms,''
\newblock Tech. {R}ep. CRIM-06/08-13, CRIM, Montreal, 2005.

\bibitem{jfa}
P.~Kenny, G.~Boulianne, P.~Ouellet, and P.~Dumouchel,
\newblock ``Joint factor analysis versus eigenchannels in speaker
  recognition,''
\newblock {\em IEEE Transactions on Audio, Speech, and Language Processing},
  vol. 15, no. 4, pp. 1435--1447, May 2007.

\bibitem{dehak2011front}
N.~Dehak, P.~Kenny, R.~Dehak, P.~Dumouchel, and P.~Ouellet,
\newblock ``Front-end factor analysis for speaker verification,''
\newblock {\em IEEE Transactions on Audio, Speech, and Language Processing},
  vol. 19, no. 4, pp. 788--798, 2011.

\bibitem{Lee2017}
K.~A. Lee and SRE’16~I4U Group,
\newblock ``The {I4U} mega fusion and collaboration for {NIST} speaker
  recognition evaluation 2016,''
\newblock in {\em Proc. Interspeech 2017}, 2017, pp. 1328--1332.

\bibitem{Torres2017}
P.~A. Torres-Carrasquillo, F.~Richardson, S.~Nercessian, D.~Sturim,
  W.~Campbell, Y.~Gwon, S.~Vattam, N.~Dehak, H.~Mallidi, P.~S. Nidadavolu,
  R.~Li, and R.~Dehak,
\newblock ``The {MIT-LL}, {JHU} and {LRDE} {NIST} 2016 speaker recognition
  evaluation system,''
\newblock in {\em Proc. Interspeech 2017}, 2017, pp. 1333--1337.

\bibitem{iitg_indigo}
N.~Kumar, R.~K. Das, S.~Jelil, B.~K. Dhanush, H.~Kashyap, K.~S.~R. Murty,
  S.~Ganapathy, R.~Sinha, and S.~R.~M. Prasanna,
\newblock ``{IITG}-{Indigo} system for {NIST} 2016 {SRE} challenge,''
\newblock in {\em Proc. Interspeech 2017}, 2017, pp. 2859--2863.

\bibitem{dnnASR}
Y.~Lei, N.~Scheffer, L.~Ferrer, and M.~McLaren,
\newblock ``A novel scheme for speaker recognition using a phonetically-aware
  deep neural network,''
\newblock in {\em IEEE International Conference on Acoustics, Speech and Signal
  Processing (ICASSP) 2014}, May 2014, pp. 1695--1699.

\bibitem{dnn_odyssey_2014}
{P. Kenny, V. Gupta, T. Stafylakis, P. Ouellet and J. Alam},
\newblock ``Deep neural networks for extracting baum-welch statistics for
  speaker recognition,''
\newblock in {\em Speaker Odyssey 2014}, 2014, pp. 293--298.

\bibitem{dnn_letter}
F.~Richardson, D.~Reynolds, and N.~Dehak,
\newblock ``Deep neural network approaches to speaker and language
  recognition,''
\newblock {\em IEEE Signal Processing Letters}, vol. 22, no. 10, pp.
  1671--1675, Oct 2015.

\bibitem{mitchDNN}
M.~McLaren, Y.~Lei, and L.~Ferrer,
\newblock ``Advances in deep neural network approaches to speaker
  recognition,''
\newblock in {\em IEEE International Conference on Acoustics, Speech and Signal
  Processing (ICASSP) 2015}, April 2015, pp. 4814--4818.

\bibitem{Variani2014Deep}
E.~Variani, X.~Lei, E.~Mcdermott, I.~L. Moreno, and J.~Gonzalez-Dominguez,
\newblock ``Deep neural networks for small footprint text-dependent speaker
  verification,''
\newblock in {\em IEEE International Conference on Acoustics, Speech and Signal
  Processing 2014}, May 2014, pp. 4052--4056.

\bibitem{end2endICASSP}
G.~Heigold, I.~Moreno, S.~Bengio, and N.~Shazeer,
\newblock ``End-to-end text-dependent speaker verification,''
\newblock in {\em IEEE International Conference on Acoustics, Speech and Signal
  Processing (ICASSP) 2016}, March 2016, pp. 5115--5119.

\bibitem{end2endSLT}
S.~X. Zhang, Z.~Chen, Y.~Zhao, J.~Li, and Y.~Gong,
\newblock ``End-to-end attention based text-dependent speaker verification,''
\newblock in {\em IEEE Spoken Language Technology Workshop (SLT) 2016}, Dec
  2016, pp. 171--178.

\bibitem{emdend2endSLT}
D.~Snyder, P.~Ghahremani, D.~Povey, D.~Garcia-Romero, Y.~Carmiel, and
  S.~Khudanpur,
\newblock ``Deep neural network-based speaker embeddings for end-to-end speaker
  verification,''
\newblock in {\em IEEE Spoken Language Technology Workshop (SLT) 2016}, Dec
  2016, pp. 165--170.

\bibitem{Li2017Deep}
C.~Li, X.~Ma, B.~Jiang, X.~Li, X.~Zhang, X.~Liu, Y.~Cao, A.~Kannan, and Z.~Zhu,
\newblock ``Deep speaker: an end-to-end neural speaker embedding system,''
\newblock {\em arXiv:1705.02304 [cs.CL]}, 2017.

\bibitem{Snyder2017embedding}
D.~Snyder, D.~Garcia-Romero, D.~Povey, and S.~Khudanpur,
\newblock ``Deep neural network embeddings for text-independent speaker
  verification,''
\newblock in {\em INTERSPEECH}, 2017, pp. 999--1003.

\bibitem{Brummer2018}
N.~Brummer, A.~Silnova, L.~Burget, and T.~Stafylakis,
\newblock ``Gaussian meta-embeddings for efficient scoring of a heavy-tailed
  plda model,''
\newblock in {\em Proc. Odyssey 2018 The Speaker and Language Recognition
  Workshop}, 2018, pp. 349--356.

\bibitem{Zhang2017End}
Chunlei Zhang and Kazuhito Koishida,
\newblock ``End-to-end text-independent speaker verification with triplet loss
  on short utterances,''
\newblock in {\em Proc. Interspeech 2017}, August 2017, pp. 1487--1491.

\bibitem{snyder2018x}
D.~Snyder, D.~Garcia-Romero, G.~Sell, D.~Povey, and S.~Khudanpur,
\newblock ``X-vectors: Robust {DNN} embeddings for speaker recognition,''
\newblock in {\em IEEE International Conference on Acoustics, Speech and Signal
  Processing (ICASSP) 2018}, April 2018, pp. 5329--5333.

\bibitem{Das2016Exploring}
R.~K. Das and S.~R.~M. Prasanna,
\newblock ``Exploring different attributes of source information for speaker
  verification with limited test data,''
\newblock {\em Journal of the Acoustical Society of America}, vol. 140, no. 1,
  pp. 184, 2016.

\bibitem{Sargin2006Multimodal}
M.~E. Sargin, E.~Erzin, Y.~Yemez, and A.~M. Tekalp,
\newblock ``Multimodal speaker identification using canonical correlation
  analysis,''
\newblock in {\em IEEE International Conference on Acoustics, Speech and Signal
  Processing, 2006. ICASSP 2006 Proceedings}, 2006, pp. I--I.

\bibitem{lt2018}
L.~Xu, K.~A. Lee, H.~Li, and Z.~Yang,
\newblock ``Co-whitening of i-vectors for short and long duration speaker
  verification,''
\newblock in {\em INTERSPEECH 2018}, September 2018, pp. 1066--1070.

\bibitem{Waibel1989}
A.~Waibel, T.~Hanazawa, G.~Hinton, K.~Shikano, and K.~J. Lang,
\newblock ``Phoneme recognition using time-delay neural networks,''
\newblock {\em IEEE Transactions on Acoustics, Speech, and Signal Processing},
  vol. 37, no. 3, pp. 328--339, Mar 1989.

\bibitem{Ioffe2015}
S.~Ioffe and C.~Szegedy,
\newblock ``Batch normalization: Accelerating deep network training by reducing
  internal covariate shift,''
\newblock in {\em Proc. ICML}, 2015, pp. 448–--456.

\bibitem{Maaten2008Visualizing}
L~Maaten and G~Hinton,
\newblock ``Visualizing data using t-sne,''
\newblock {\em Journal of Machine Learning Research}, vol. 9, no. 2605, pp.
  2579--2605, 2008.

\bibitem{nist2010}
``The {NIST} year 2010 speaker recognition evaluation plan,''
\newblock April 2010.

\bibitem{prince2007probabilistic}
S.~J.~D. Prince and J.~H. Elder,
\newblock ``Probabilistic linear discriminant analysis for inferences about
  identity,''
\newblock in {\em Computer Vision, 2007. ICCV 2007. IEEE 11th International
  Conference on}. IEEE, 2007, pp. 1--8.

\bibitem{Povey_ASRU2011}
D.~Povey, A.~Ghoshal, G.~Boulianne, L.~Burget, O.~Glembek, N.~Goel,
  M.~Hannemann, P.~Motlicek, Y.~Qian, P.~Schwarz, J.~Silovsky, G.~Stemmer, and
  K.~Vesely,
\newblock ``The kaldi speech recognition toolkit,''
\newblock in {\em IEEE Workshop on Automatic Speech Recognition and
  Understanding 2011}, Dec. 2011.

\bibitem{Mclaren2018}
M.~Mclaren, D.~Castan, M.~K. Nandwana, L.~Ferrer, and E.~Y{\i}lmaz,
\newblock ``How to train your speaker embeddings extractor,''
\newblock in {\em Proc. Odyssey 2018 The Speaker and Language Recognition
  Workshop}, 2018, pp. 327--334.

\bibitem{Novoselov2018}
S.~Novoselov, A.~Shulipa, I.~Kremnev, A.~Kozlov, and V.~Shchemelinin,
\newblock ``On deep speaker embeddings for text-independent speaker
  recognition,''
\newblock in {\em Proc. Odyssey 2018 The Speaker and Language Recognition
  Workshop}, 2018, pp. 378--385.

\end{thebibliography}

\end{document}